\documentclass[%
 reprint,
 amsmath,amssymb,
 aps,
]{revtex4-2}

\usepackage{graphicx}
\usepackage{dcolumn}
\usepackage{bm}
\usepackage{siunitx}
\usepackage{color}
\usepackage{autobreak}
\usepackage{braket}
\usepackage{here}
\usepackage{array}
\usepackage{appendix}
\usepackage[normalem]{ulem}

\begin{document}
\preprint{APS/123-QED}

\title{Observation of nonreciprocal diffraction
of surface acoustic wave} 

\author{Y. Nii$^{1}$}
\email{yoichi.nii.c1@tohoku.ac.jp}
\author{K. Yamamoto$^{2,3}$}
\author{M. Kanno$^{1}$}
\author{S. Maekawa$^{3,2,4}$}
\author{Y. Onose$^{1}$}

\affiliation{
$^{1}$Institute for Materials Research, Tohoku University, Sendai 980-8577, Japan.\\
$^{2}$Advanced Science Research Center, Japan Atomic Energy Agency, Tokai 319-1195, Japan\\
$^{3}$Center for Emergent Matter Science (CEMS), RIKEN , Wako 351-0198, Japan.
$^{4}$Kavli Institute for Theoretical Sciences, University of Chinese Academy of Sciences , Beijing 100190,  China
}

\begin{abstract}
Rectification phenomenon caused by the simultaneous breaking of time reversal and spatial inversion symmetries has been extended to a wide range of (quasi)particles and waves; however, the nonreciprocal diffraction, which is the imbalance of upward and downward deflections, was previously observed only for photons and remained to be extended to other (quasi)particles. In this study, we present evidence of the nonreciprocal diffraction of surface acoustic wave (SAW) utilizing a magnetoelastic grating on a SAW device. Asymmetric diffraction intensities were observed when the ferromagnetic resonance was acoustically excited. Based on a theoretical model, we attribute the microscopic origin of this phenomenon to the resonant scattering involving ferromagnetic resonance excitations. The novel property may pave an avenue to further development of SAW devices for various purposes, including microwave communications and quantum engineering applications.
\end{abstract}

\maketitle
In systems with broken time-reversal (TRS) and spatial inversion symmetries (SIS), directional anisotropy emerges among (quasi)particles and waves propagating in forward and backward directions \cite{tokura}. The electrical conductance or wave-transmission amplitude varies upon reversing the transport (propagation) direction, which is referred to as nonreciprocal directional response. Initially demonstrated in optical (photonic) properties \cite{hopfield, rikkenMCH, rikkenME}, this phenomenon has been extended to electrical transport \cite{rikkenE, ideue}, as well as magnon \cite{iguchi, seki} and phonon transmission \cite{heil, sasaki, xu, kuss, shah, nomura, sasaki2021}. The preferred direction of propagation is reversed by either TRS or SIS operation and can be controlled by the external fields. This unidirectional nature has been attracting significant attention \cite{tokura, nassar, cheong}, both in terms of potential applications and fundamental science. There is a related but distinct class of phenomena denoted as nonreciprocal diffraction, in which diffraction intensity of certain waves depends on the sign of scattering vector $\Delta \bm{Q}=\bm{q}-\bm{k}$ ($\bm{k}$ and $\bm{q}$ are wave vectors at the initial and final states, respectively). The nonreciprocal diffraction has been demonstrated in optics \cite{kida, kida2}, but its extension to other waves remains unexplored.\par

\begin{figure}
    \centering
    \includegraphics[width=3.375in]{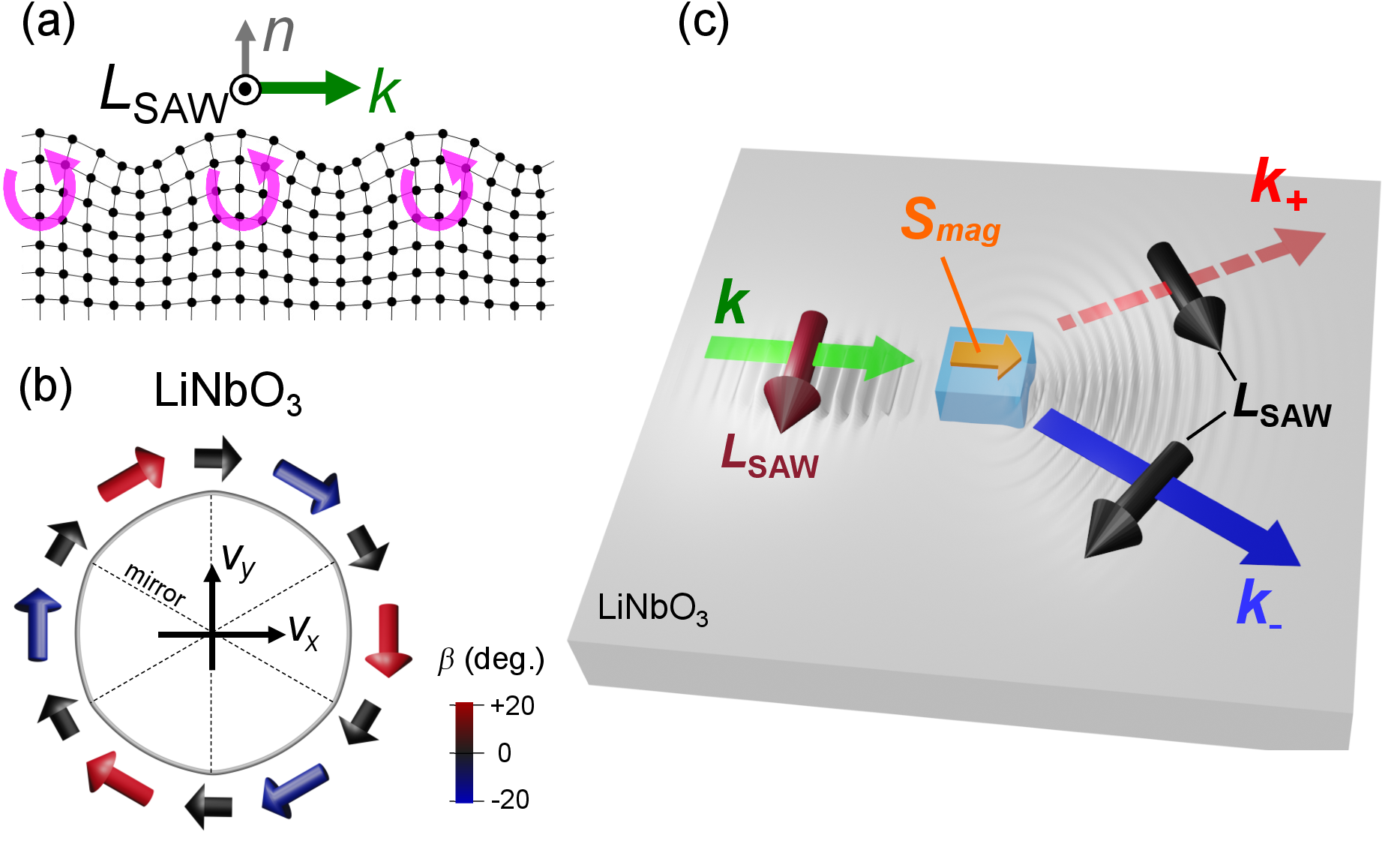}
    \caption{(a) Cross sectional view of Rayleigh-type SAW. $\bm{L}_{\rm{SAW}}$ represents angular momentum carried by SAW. The purple arrows represent elliptical rotation of the lattice points. The direction of $\bm{L}_{\rm{SAW}}$ is nearly perpendicular to both the surface normal ($\bm{n}$) and $\bm{k}$. (b) Calculated $\bm{L}_{\rm{SAW}}$ distribution superimposed on the velocity curve of LiNbO$_3$ in the Z-plane. The arrows denote $\bm{L}_{\rm{SAW}}$, and their colors indicate the canting angle $\beta$ toward the out-of-plane direction. The calculation was performed using a finite element method by COMSOL Multiphysics 5.6 as described in the Supplemental material \cite{YNii_supple}. (c) Illustration of nonreciprocal scattering of SAW. Due to the relation between $\bm{S}_{\rm{mag}}$ and $\bm{L}_{\rm{SAW}}$ of scattered SAW, the amplitudes of upward and downward scattering differ from each other.}
    \label{fig1}
\end{figure}

In this study, we examine the nonreciprocal diffraction of surface acoustic wave (SAW). As illustrated in Fig. 1(a), SAW is an elastic wave localized on the surface of solids. The fundamental properties of SAW are established and widely recognized \cite{landau, bauer}. In contrast to conventional bulk acoustic waves, in which the longitudinal and transverse modes are distinct from each other, Rayleigh-type SAW is described by a superposition of these two polarizations. Consequently, SAW exhibits an ellipic motion of material points near the surface [Fig. 1(a)]. In an isotropic continuum medium, the plane of elliptic oscillation is spanned by the surface normal $\bm{n}$ and the SAW momentum $\bm{k}$ as described in a standard textbook \cite{landau}.  Because the acoustic wave is a low-energy phonon mode, 
the elliptic oscillation can be characterized by the angular momenum of phonon defined by Zhang and Niu \cite{PAM}.  The angular momentum of SAW ($\bm{L}_{\rm{SAW}}$) is proportional to the vector product of $\bm{k}$ and $\bm{n}$ in an isotropic continuum medium. Figure 1(b) shows $\bm{L}_{\rm{SAW}}$ in $\bm{k}$-space calculated for the actual LiNbO$_3$ crystal structure \cite{YNii_supple}. Although there is a minor out-of-plane component of $\bm{L}_{\rm{SAW}}$ in the actual anisotropic crystal, the fundamental trend remains consistent with the relation $\bm{L}_{\rm{SAW}}\propto \bm{k} \times \bm{n}$. Noted that the $\bm{k}$-dependent $\bm{L}_{\rm{SAW}}$ is an analog of the $\bm{k}$-dependent spin angular momentum in electronic surface states or other symmetry-broken systems \cite{SML}. \par
Diffraction of SAW by a ferromagnetic object with a spin angular momentum $\bm{S}_{\rm{mag}}$ should become nonreciprocal; the intensities of upward $\bm{k}_+$ and downward $\bm{k}_-$ scattered waves become distinct from each other [Fig. 1(c)]. When $\bm{S}_{\rm{mag}}$ is parallel to the incident wave vector of SAW, $\bm{L}_{\rm{SAW}}\cdot \bm{S}_{\rm{mag}} < 0$ for the $\bm{k}_+$ wave, whereas $\bm{L}_{\rm{SAW}}\cdot \bm{S}_{\rm{mag}} > 0$ for the $\bm{k}_-$ wave, considering the $\bm{k}$-dependence of $\bm{L}_{\rm{SAW}}$ [Fig. 1(b)]. Therefore, the properties of these two scattered SAWs should be inequivalent, suggesting the emergence of nonreciprocity in the scattering amplitude. The asymmetry of scattering can be reversed by the reversal of $\bm{S}_{\rm{mag}}$ because it depends on the sign of $\bm{L}_{\rm{SAW}}\cdot \bm{S}_{\rm{mag}}$. Motivated by these symmetry arguments, we experimentally demonstrated the nonreciprocal diffraction of SAW by employing a periodic array of magnetoelastic nanowires (i .e., SAW grating) for the enhancement of diffraction. In addition, we discussed both the qualitative and quantitative aspects of the nonreciprocal diffraction based on a theoretical model of the resonant scattering of SAW.

\begin{figure}
    \centering
    \includegraphics[width=3.375in]{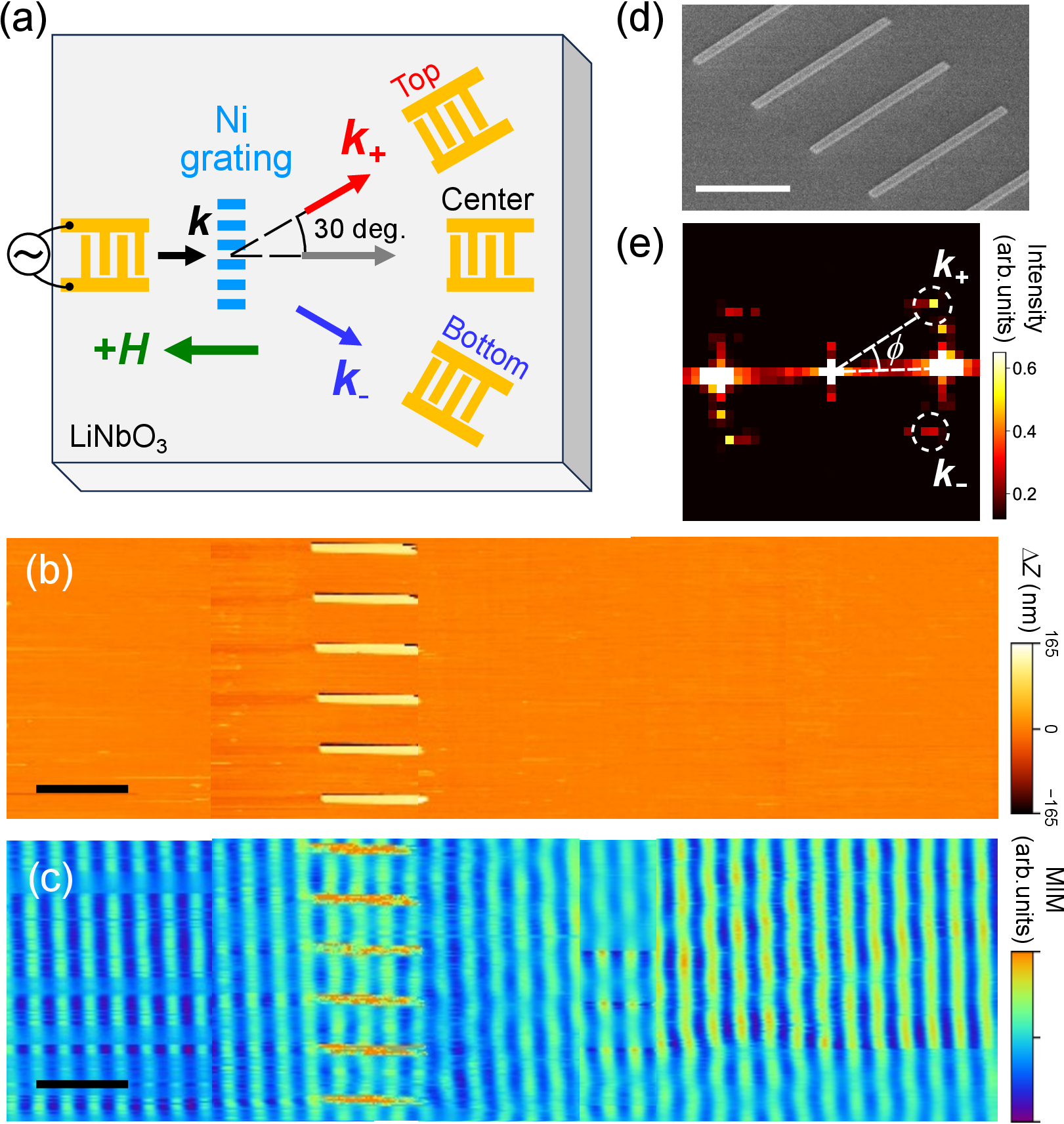}
    \caption{(a) Experimental set-up for nonreciprocal diffraction measurement of SAW. The left IDT launches SAW to the Ni grating. Transmitted and diffracted SAWs were detected by three IDTs on the right denoted as center, top, and bottom IDTs, respectively. 
    (b) AFM and (c) SAW images around the Ni grating, which were obtained simultaneously using pulse-modulated MIM. During the scanning the cantilever tip over the surface of the device, SAW at 2.6 GHz was launched from the left IDT to right. Scale bars are 5 $\mu$m. (d) Scanning electron microscopy (SEM) image of the Ni grating. Scale bar is 3 $\mu$m. (e) FT analysis of the MIM image. The region on the right side of the MIM image was used for this FT analysis. Note that the MIM image shown in (c) is a snapshot and does not include time dependence information; thus, FT image is symmetric with respect to the origin. All the AFM, MIM, and SEM images were taken without magnetic field at room temperature.
    }
    \label{fig2}
\end{figure}

Figure 2(a) shows the SAW device for the nonreciprocal SAW diffraction. It consists of ferromagnetic Ni grating [Fig. 2(d)] and four interdigital transducers (IDT) on Z-cut LiNbO$_3$ piezoelectric substrate. The left IDT excites SAW, and the right-center IDT receives the transmitted SAW. The right-top and right-bottom receive the diffracted SAWs. The grating structure was utilized to enhance the scattering intensity owing to constructive interference at the Bragg condition. The grating and the IDTs were fabricated by electron beam lithography followed by electron beam evaporation of Ni and Ti/Au, respectively. The Ni nanowires have a thickness, width, and length of 110 nm, 300 nm, and 5.3 $\mu$m, respectively. The grating period is $d$ = 2.71 $\mu$m and the wavelength of SAW is $\lambda_{\rm{SAW}}$ = 1.44 $\mu$m at 2.6 GHz \cite{SAW_velocity}, leading to Bragg diffraction angle of $\phi$ = ± 32 deg according to Bragg’s law $d\sin\phi$ = $n\lambda_{\rm{SAW}}$ ($n$ represents the diffraction order, assumed to be ± 1). To verify this relationship, we visualized SAW propagation through the Ni grating using scanning microwave impedance microscopy (MIM) \cite{SAW_imaging}. This technique can be applied to probe intriguing gigahertz sounds, such as topological sound \cite{zhang, nii}, interference \cite{zheng}, and acoustic cavity mode \cite{AcousticResonator} with sub-$\mu$m spatial resolution. For MIM imaging, a similar device was separately fabricated on the LiNbO$_3$ substrate. Details regarding MIM and device configurations are provided in the Supplemental material \cite{YNii_supple}. Figure 2(c) shows MIM image corresponding to SAW propagating across the Ni grating. For comparison, the atomic force microscopy (AFM) image is shown in Fig. 2(b). A horizontal periodic change in contrast on the left side of MIM image indicates the plane wave of SAW. The SAW mostly transmits through the Ni grating, but is partially diffracted by it. The transmitted and diffracted SAWs show an interference pattern on the right side; the SAW amplitude is also modulated in the vertical direction. Fourier transformation (FT) of the interference image yields bright spots at diffraction angles of $\phi \simeq \pm$ 31 deg [Fig. 2(e)], which is close to the calculated value shown above. Based on this measurement, we designed the tilting angles of the two IDTs for receiving the diffracted SAWs as $\pm$ 30 deg, which is equivalent to one of the principal axes ($Y$-axis) of LiNbO$_3$. In the actual device, the IDT fingers have a length of 30 $\mu$m, allowing the detection of diffracted SAWs ranging from $\phi$ = 26 to 34 deg \cite {YNii_supple}, covering the relevant range in the present study.

\begin{figure}
    \centering
    \includegraphics[width=2.5in]{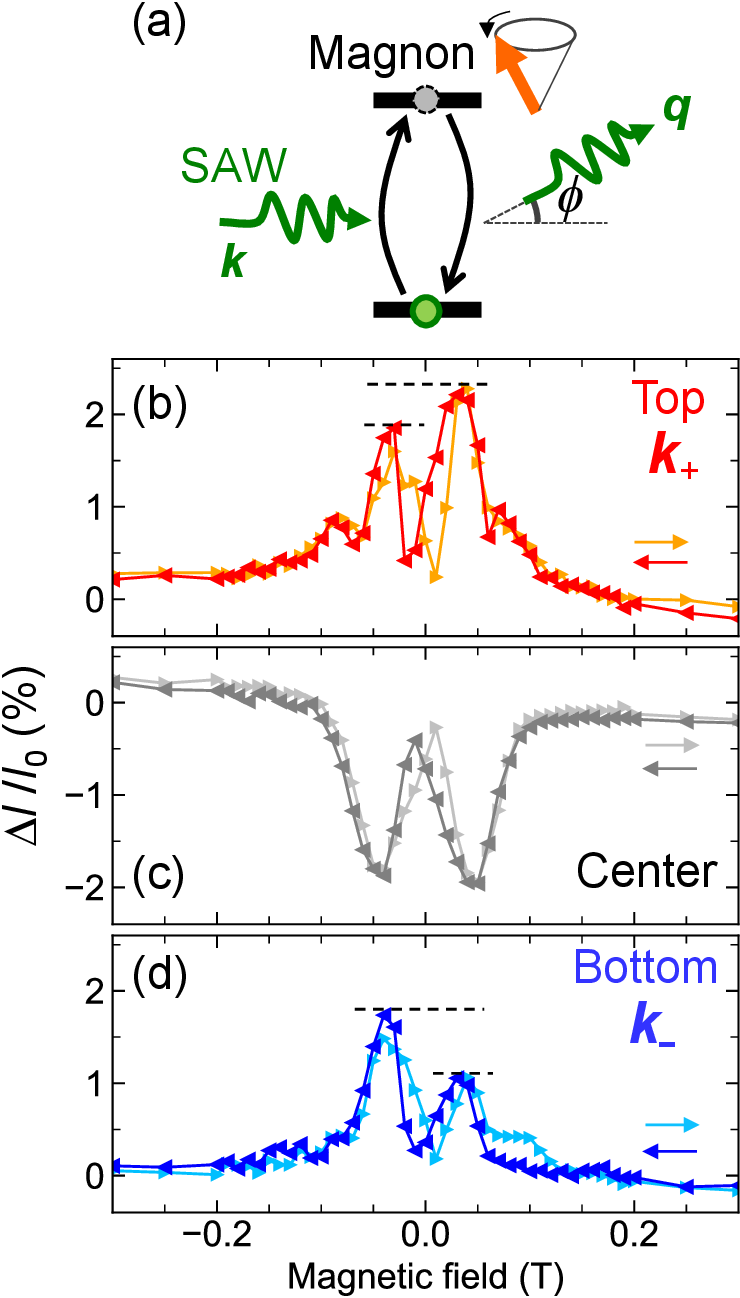}
    \caption{(a) Resonant SAW scattering process via ferromagnetic resonance (magnon) excitation. In the first quantum process, an incident SAW excites magnon. In the subsequent process, the excited magnon emits SAW with various angles $\phi$. (b)-(d) Magnetic field dependence of the relative change in the SAW amplitude measured by (b) top ($\phi$ = 30 deg), (c) center ($\phi$ = 0 deg), and (d) bottom ($\phi$ = $-$30 deg) IDTs, respectively. While the transmitted SAW was measured at the center IDT, the upward and downward diffracted SAW intensities were measured by the top and bottom IDTs, respectively.
    }
    \label{fig1}
\end{figure}

\par
Next, we discuss the magnetic field dependence of transmitted and diffracted SAW intensities, which are quantitatively evaluated by microwave power transmittances $|\widetilde{S_{21}}|$ from the left IDT to one of the right IDTs. $|\widetilde{S_{21}}|$ is $S$-parameter measured with a vector network analyzer (Agilent E5071C) averaged in the frequency range from 2.60 GHz to 2.64 GHz after a time-domain gating analysis (details are shown in the Supplemental materials \cite{YNii_supple}). All the subsequent measurements were conducted at 260 K in a split-type superconducting magnet with a rotation stage. This temperature was selected for its superior temperature stability in our measurement system. At first, the magnetic field is applied along the $\bm{k}$ direction. Here, the positive (negative) sign of the magnetic field is defined as antiparallel (parallel) to the $\bm{k}$ of the incident SAW. Figure 3(c) shows the relative change in $|\widetilde{S_{21}}|$ with respect to the magnetic field ($H$), expressed as $\Delta I/I_0 = [|\widetilde{S_{21}}(H)|-|\widetilde{S_{21}} (H_{ref})|]/|\widetilde{S_{21}}(H_{ref})|$. This was measured between the left and right-center IDTs, probing transmission of SAW. Here, $I_0 = |\widetilde{S_{21}}(H_{ref})|$ is $|\widetilde{S_{21}}|$ at high magnetic fields, $\mu_0 H_{\rm{ref}}$ = $\pm$ 300 mT, where the ferromagnetic magnon frequency is considerably higher than the SAW frequency, and the SAW and magnon modes are decoupled \cite{weiler, sasaki}. At the low magnetic fields, two dips in $\Delta I/I_0$ are apparent, one occurring around + 50 mT and the other around $-$50 mT. Similar dips in SAW transmittance (or peaks in absorption) were commonly reported in previous studies \cite{weiler, Dreher, sasaki, kuss, xu, shah}, which were attributed to absorption by so-called acoustically-induced ferromagnetic resonance (acoustic-FMR). As SAW traverses ferromagnetic materials, magnetoelastic coupling produces torque on the magnetic moment \cite{kittel}, which transfers the energy from SAW to FMR when their frequencies coincide. The nearly symmetric depths of these two dips indicate minimal nonreciprocal directional propagation for the $\bm{k}$ $||$ $\bm{H}$ configuration, consistent with the previous studies \cite{sasaki, xu, kuss, shah}.

\begin{figure}
    \centering
    \includegraphics[width=3.375in]{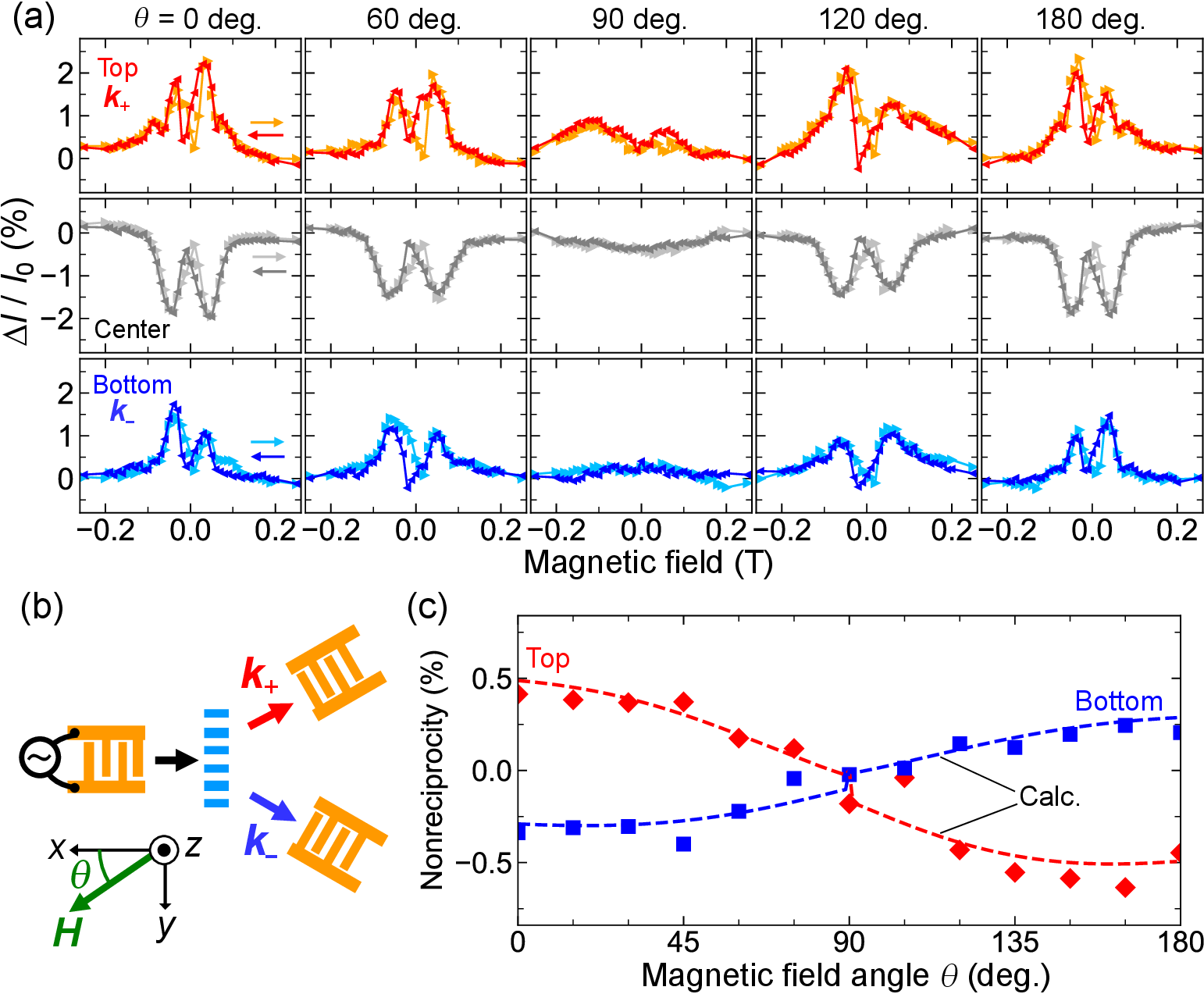}
    \caption{(a). Magnetic field dependence of $\Delta I/I_0$ for the transmitted and the diffracted ($\bm{k}_{\pm}$) SAWs at various magnetic field angles $\theta$. (b) Definition of magnetic field angle $\theta$ with respect to SAW device. (c) Nonreciprocity of diffracted SAWs as a function of magnetic field angle. The nonreciprocity is evaluated by the difference of peak intensities at positive and magnetic fields (see text). Dashed lines are the result of fitting by the theoretically obtained nonreciprocity \cite{YNii_supple}.
    }
    \label{fig1}
\end{figure}
\par
Let us move on to the SAW diffraction. Figures 3(b) and 3(d) show $\Delta I/I_0$ for the upward ($\bm{k}_+$) and downward ($\bm{k}_-$) diffracted SAWs, respectively. While large field-independent component of diffraction was observed \cite{YNii_supple}, $\Delta I/I_0$  exhibited a sharp increase of approximately 2 $\%$ at magnetic fields corresponding to the two dips in SAW transmittance [Fig. 3(c)]. Therefore, this enhancement suggests that the resonant scattering process involving a magnetic excitation shown in Fig. 3(a) is responsible for the observed magnetic field dependence. More importantly, these peak heights were dependent on the polarity of magnetic fields, as emphasized by the dashed lines. This asymmetric field dependence is reversed by the reversal of diffraction direction as shown in Fig. 3(d). These features are in agreement with the nonreciprocal diffraction.
\par
To further study the origin of nonreciprocal SAW diffraction, we investigated magnetic field angle dependence in the configuration shown in Fig. 4(b). Figure 4(a) represents the magnetic field dependence of SAW diffraction intensities at various field angles ($\theta$) in comparison with SAW transmission intensities. With an increase in the angle from $\theta$ = 0 deg, the asymmetry of diffracted intensities upon the magnetic field reversal gradually diminishes. At $\theta$ = 90 deg, the diffraction intensity is nearly constant and two dip structures in transmittance disappear. These are attributable to the suppression of acoustic FMR excitation in this orthogonal magnetic field angle as reported in the literatures \cite{sasaki,xu,kuss, shah}. For the larger magnetic field angles, the asymmetry of diffracted intensities in field polarity reverses.
At any field angle, the asymmetries for the upward and downward diffractions are opposite to each other. Figure 4(c) shows the angular dependence of $[I_{\rm{max}}(+H) – I_{\rm{max}}(-H)]/I_0$, where $I_{\rm{max}}(\pm H)$ is the average of ten data points around the maximal value in the positive or negative magnetic field region, respectively. For both diffraction directions, $[I_{\rm{max}}(+H) – I_{\rm{max}}(-H)]/I_0$ decreases with the magnetic field angle and shows a sign change at $\theta$ = 90 deg.
\par
Having experimentally established the nonreciprocal SAW diffraction and its magnetic field dependence, we now interpret these observations in terms of the resonant scattering of SAW by magnetic excitations. As mentioned above, the magnetic field dependence suggests the origin of nonreciprocal SAW diffraction should be ascribed to the acoustically driven FMR as shown in Fig. 3(a). 
To capture the essential physics without overcomplicating the analysis, we theoretically model the diffraction as a quantum mechanical scattering process off the spin degrees of freedom $\bm{S}_{\rm{mag}}$. When the SAW and spin interact through the Hamiltonian of magnetoelastic interaction $\mathcal{H}_{me}$, the scattering amplitude is given by
\begin{align}
\sum_{\bm{\alpha}}\displaystyle\frac{\braket{\bm{q}|\mathcal{H}_{me}|\bm{\alpha}}\braket{\bm{\alpha}|\mathcal{H}_{me}|\bm{k}}}{\omega_\alpha-\omega_k+i\Gamma},
\end{align}
where $\ket{\bm{k}}$, $\ket{\bm{q}}$, $\ket{\bm{\alpha}}$ are the initial and scattered states of SAW, and an FMR excited intermediate state of spin, and $\omega_{\bm{\alpha}}$, $\omega_{\bm{k}}$, $\Gamma$ are frequencies of magnon (FMR) and SAW, and lifetime of the intermediate state, respectively. In this formula, the matrix element $\braket{\bm{q} |\mathcal{H}_{me}|\bm{\alpha}}$ gives rise to the nonreciprocity, ${\it i.e.}$, inequivalence of $\ket{\bm{q}} = \ket{\bm{k}_+}$ and $\ket{\bm{q}} = \ket{\bm{k}_-}$ in the scattering process owing to a contribution proportional to $\bm{L}_{\rm{SAW}}\cdot \bm{S}_{\rm{mag}}$. While the angular momentum of the magnon $\ket{\bm{\alpha}}$ state is antiparallel to the ferromagnetic moment $\bm{S}_{\rm{mag}}$, $\bm{L}_{\rm{SAW}}$ of scattered SAW ($\ket{\bm{q}}$)  depends on the scattering direction as shown in Fig. 1(b). When the magnetic field is at $\theta$ = 0 deg, $\bm{L}_{\rm{SAW}}\cdot \bm{S}_{\rm{mag}} < 0$ for positive scattering angle and $\bm{L}_{\rm{SAW}}\cdot \bm{S}_{\rm{mag}} > 0$ for negative scattering angle as mentioned previously. Therefore, this matrix element should account for the diffraction asymmetry between the upward and downward scatterings. For a more quantitative argument, we set up a particular model Hamiltonian assuming isotropic elastic properties and magnetoelastic interactions, and computed the magnetic field angle dependence of the scattering amplitude. The detail of calculation is presented in the supplemental material \cite{YNii_supple}. The calculated field angle dependence of the nonreciprocal diffraction for the top and bottom IDTs are plotted in Fig. 4(c) alongside the experimental data points. The agreement is satisfactory and supports our inference concerning the origin of the observed nonreciprocal SAW diffraction.\par
Previously, it has been reported that systems with modulation in both space and time exhibit various types of acoustic nonreciprocity \cite{STM1,STM2,STM3,STM4}. In their systems, TRS was broken by temporally modulated external fields, giving rise to nonreciprocal shifts in $\bm{k}$-direction, frequency, and intensity of acoustic waves. In contrast, our system intrinsically breaks TRS in ferromagnets and does not require temporal modulation. As a result of resonant scattering utilized in the present study, the nonreciprocity manifests itself not in frequency or wavenumber but in the diffraction intensity.\par
In conclusion, we investigated SAW propagation through the ferromagnetic Ni grating device on LiNbO$_3$ substrate. Our study demonstrated nonreciprocal diffraction of SAW, a phenomenon fundamentally distinct from the previous observations of nonreciprocal propagation of SAW \cite{heil, sasaki,sasaki2021, xu, kuss, shah}. The theoretical calculation unveiled that the resonant scattering process involving ferromagnetic resonance is responsible for the nonreciprocal SAW diffraction. The result clearly shows that the phenomenon of nonreciprocal diffraction is not restricted to the optical property but can be extended to other wave-like excitations, similar to the case of nonreciprocal directional response. 
The present study presents a novel SAW functionality arising from artificial patterning of magnetoelastic media as reported recently \cite{Liyang}, which contributes to the further development of SAW-based technologies \cite{delsing} such as microwave communications \cite{morgan}, sensing \cite{SAWsensor}, and quantum state engineering \cite{QuantumSAW}.


\section*{Acknowledgement}
We are grateful to T. Seki for technical advice for fabricating device using electron beam lithography. This work is supported by JSPS KAKENHI (Grants No. JP20K03828, JP21H01036, JP22H04461, 21K13886, 24K01283, 24H00189, 24H01638, 24K00576), PRESTO (Grant No. JPMJPR19L6, JPMJPR20LB), FOREST (Grant No. JPMJFR2133), and Murata Science Foundation.

\bibliography{SAW}
\end{document}


\preprint{APS/123-QED}

\title{Supplemental Material for "Observation of nonreciprocal diffraction
of surface acoustic wave"} 

\author{Y. Nii$^{1}$}
\email{yoichi.nii.c1@tohoku.ac.jp}
\author{K. Yamamoto$^{2,3}$}
\author{M. Kanno$^{1}$}
\author{S. Maekawa$^{3,2,4}$}
\author{Y. Onose$^{1}$}

\affiliation{
$^{1}$Institute for Materials Research, Tohoku University, Sendai 980-8577, Japan.\\
$^{2}$Advanced Science Research Center, Japan Atomic Energy Agency, Tokai 319-1195, Japan\\
$^{3}$Center for Emergent Matter Science (CEMS), RIKEN , Wako 351-0198, Japan.\\
$^{4}$Kavli Institute for Theoretical Sciences, University of Chinese Academy of Sciences , Beijing 100190,  China
}

\maketitle
\subsection{Device fabrication}

Figures S1(a)-S1(c) show a fabricated device on $Z$-cut LiNbO$_3$. In the initial step, the Ni grating was prepared by the standard electron beam lithography technique. A grating pattern was drawn by electron beam followed by the development. Subsequently, Ti (5 nm) and Ni (110 nm) were evaporated using an electron beam, and the lift-off technique was used to obtain the grating pattern. In the subsequent step, interdigital transducers (IDTs) were fabricated using a similar procedure, but Ti (5 nm) and Au (60 nm) were evaporated by electron beam for the case of IDTs. These IDTs have a chirped structure, with the period continuously varying from 1.36 $\mu$m to 1.41 $\mu$m.
The propagation direction of the incident surface acoustic wave (SAW) from the left IDT is designed to be parallel to the $X$-axis of $Z$-cut LiNbO$_3$, while the right-top and the right-bottom IDTs are tilted by $\pm$ 30 deg. The propagation direction of the diffracted SAWs along $\pm$ 30 deg is parallel to the $Y$-axis of LiNbO$_3$. The length of fingers for these straight and tilted IDTs are 90 $\mu$m and 30 $\mu$m, respectively. 
The distance between the grating and the tilted IDT is 200 $\mu$m, and thus the tilted IDT covers angular ranges of $\Delta \phi = \tan^{-1}(30/200) \simeq $ 8.5 deg. Then, diffracted SAWs of $\phi = $ 30 $\pm$ 4.25 deg can be detectable. For the microwave impedance microscopy (MIM) experiment, a similar device was fabricated, in which there exists only center IDTs to excite SAW.

\begin{figure}
    \centering
    \includegraphics[width=6.7in]{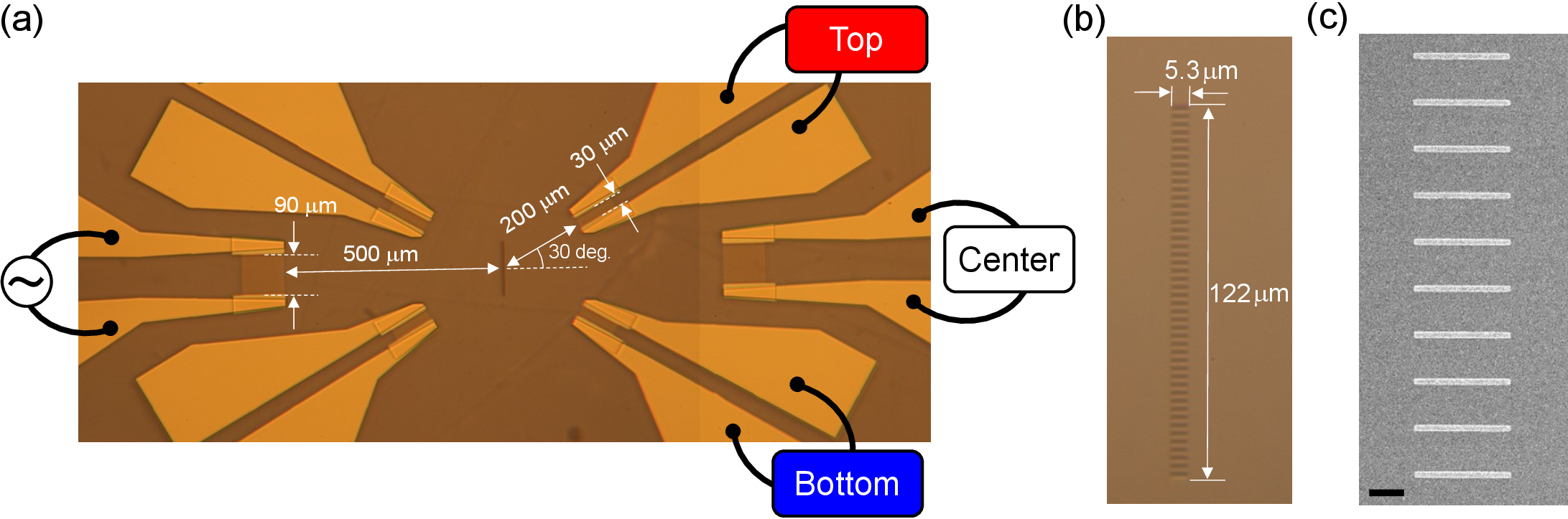}
    \caption{(a) Optical micrographs of a SAW device used for the magnetic field dependence of nonreciprocal diffraction measurement. (b) Ni grating on the center of the SAW device.(c) Scanning electron microscopy (SEM) image of the Ni grating. Scale bar is 2 $\mu$m.}
    \label{fig1}
\end{figure}

\subsection{Measurement and analysis of S$_{21}$}

 All the $S_{21}$ measurements were performed by using a vector network analyzer (E5071C, Agilent) with a home-built probe inserted into a superconducting magnet (SpectromagPT, Oxford Instruments). The probe has four semi-rigid coaxial cables and all the four IDTs were electrically connected to these semi-rigid cables; one for exciting SAW and the otheres (referred to as top, bottom, and center) for detection. To investigate the magnetic field angle dependence, the probe was rotated about the vertical direction in a horizontal magnetic field. Special care was taken for the temperature stability to eliminate signal drift during the measurement. We had waited at least for 1 hour after rotating the probe to obtain stability.
 For the SAW propagation measurements, the time-domain gating technique was employed, commonly used to extract SAW signals and to eliminate electrical crosstalk \cite{kobayashi, sasakiCBO, sasaki2021}. Figure S2(a) shows power transmission spectra ($S_{21}$) from the left IDT to one of the right IDTs denoted as Top, Center, and Bottom in Fig. S1(a). The orange and blue colors correspond to raw and filtered spectra using time-domain gating technique. Figure S2(b) represent Fourier transformed $S_{21}$ shown in time-domain, where the time-delayed SAW signals are clearly evident at 300-400 ns for the center IDT and at 200-300 ns for the top and bottom IDTs, respectively. The difference of these time-of-flight values is primarily attributed to the distance between the grating and the detecting IDTs on the right side [see Fig. S1(a)]. Performing an inverse Fourier transformation of these time-delayed SAW signals, filtered spectra were obtained, as shown in Fig. S2(a). Due to anisotropy in the sound velocity in $Z$-plane of LiNbO$_3$ as shown in Fig. 1(b), peak frequency in $|\widetilde{S_{21}}|$ slightly differs between the center and top (bottom) IDTs. The filtered $S_{21}$
 spectra was averaged from 2.60 GHz to 2.64 GHz after converting decibel unit of $S_{21}$ to linear power using $10^{(S_{21}/10)}$. Then, we obtained averaged transmission power represented by  $|\widetilde{S_{21}}|$. Based on this analysis, we obtained magnetic field dependence of $\Delta I/I_0 = [|\widetilde{S_{21}}(H)|-|\widetilde{S_{21}}(H_{ref})|/|\widetilde{S_{21}}(H_{ref})|$ as exhibited in Fig. S3(a)-S3(c). Here, $I_0$ = $|\widetilde{S_{21}}(H_{ref})|$ is $|\widetilde{S_{21}}|$ at high enough magnetic fields, $\mu_0H_{ref}$ = $\pm$ 300 mT, where the ferromagnetic resonance frequnecy is much higher than the SAW frequency, and the SAW and magnon modes are decoupled \cite{sasaki, Weiler}. The magnetic field direction is varied from 0 to 180 deg.

\begin{figure}
    \centering
    \includegraphics[width=6.7in]{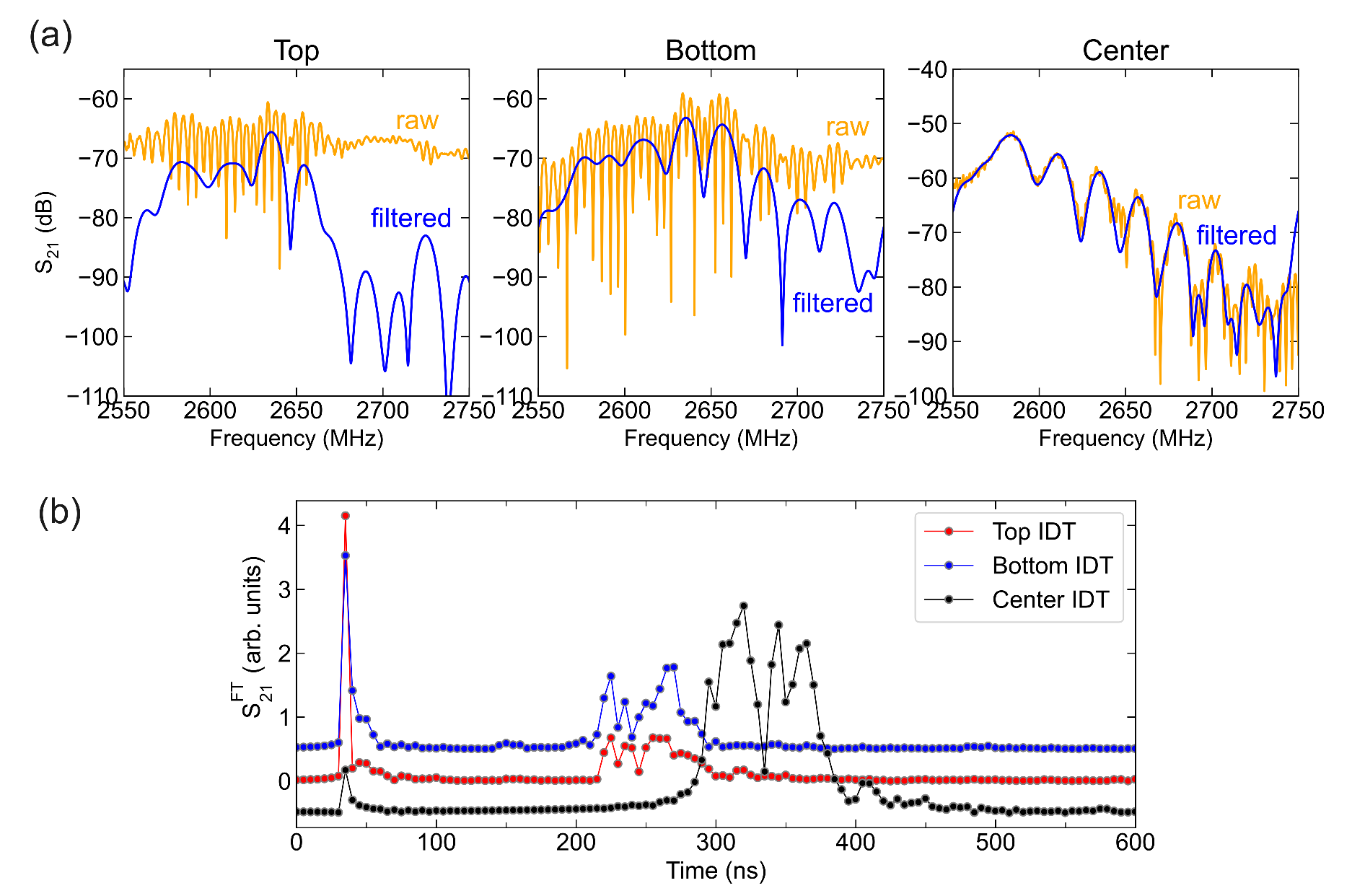}
    \caption{(a) Transmission spectra $S_{21}$ from the left IDT to one of the right IDTs denoted as top, bottom, and center. Raw and time domain filtered spectrum were shown by orange and blue colors, respectively. (b) The Fourier transformed $S_{21}$ represented in time-domain. For the filtering in the time-domain, we utilized time-delayed signals observed around 300-400 ns at center IDT and 200-300 ns at top and bottom IDTs corresponding to transmitted SAW and diffracted SAWs, respectively.
}
    \label{fig1}
\end{figure}

\begin{figure}
    \centering
    \includegraphics[width=5.5in]{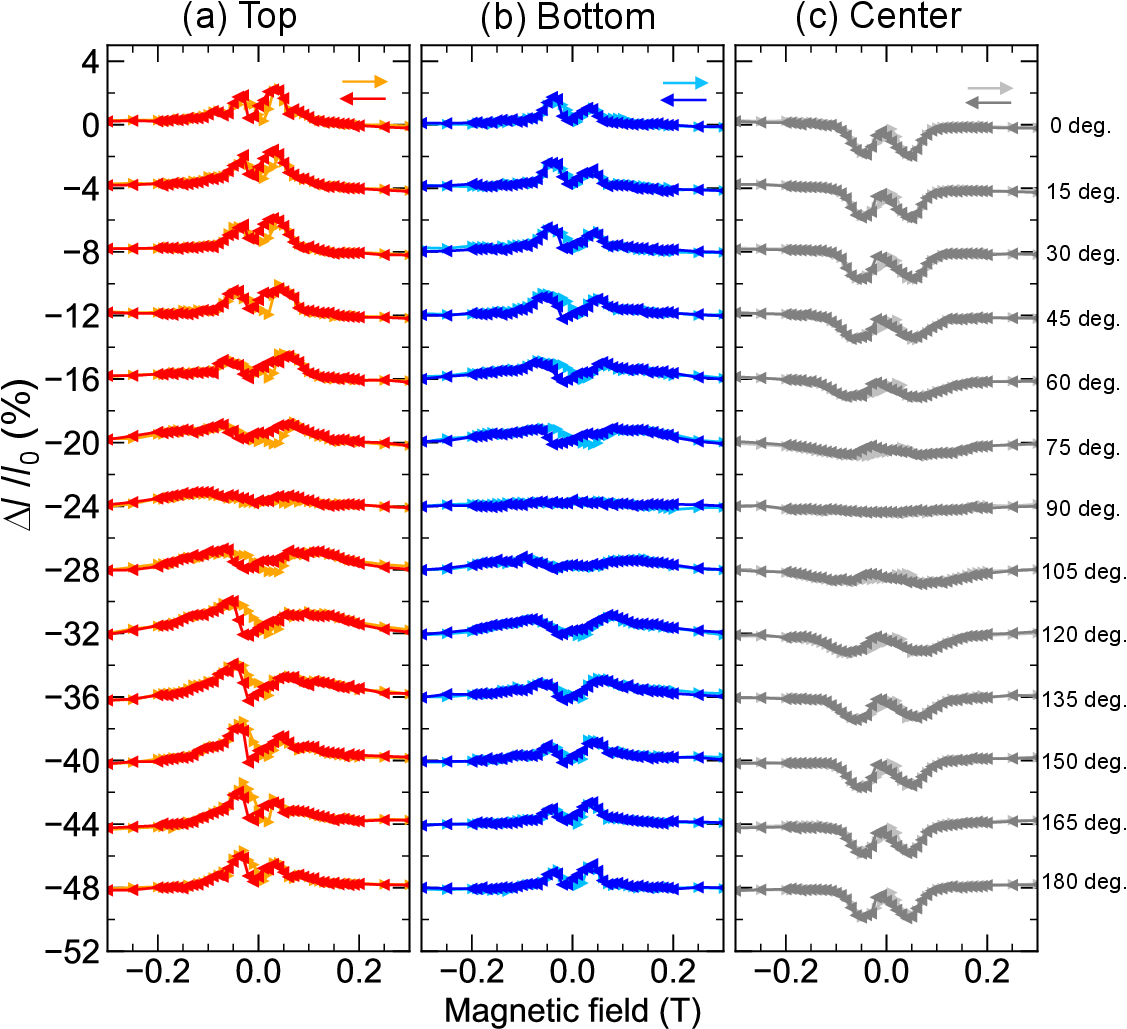}
    \caption{Magnetic field dependence of $\Delta I/I_0$ measured by (a) top ($\phi$ = + 30 deg), (b) bottom ($\phi$ = - 30 deg), and (c) center ($\phi$ = 0 deg) IDTs at various magnetic field angles $\theta$. 
    }
    \label{fig1}
\end{figure}

\subsection{SAW imaging using scanning microwave impedance microscopy}
To visualize SAW diffraction, we utilized a pulse-type MIM imaging technique \cite{nii} as schematically illustrated in Fig S4. This MIM technique is basically similar to the conventional transmission-type MIM \cite{SAW_imaging,zhang, zheng}, but the unique feature is the use of pulse-modulated microwave, in which one can separate SAW signal from electromagnetic crosstalk in time domain. The instruments and microwave components utilized in this study were the same as in a previous study \cite{nii}.
\par
In this MIM, we first apply pulse-modulated microwave voltage to IDT to excite SAW. The propagation of SAW involves spatiotemporal oscillation of electric potential via the piezoelectric property of the LiNbO$_3$ substrate. This electrical signal was locally detected by contacting a conductive cantilever directly on the $Z$-cut LiNbO$_3$, then amplified and demodulated using the IQ mixer. 
By using a Boxcar integrator, we eliminated electrical crosstalk and detected the real and imaginary part of the demodulated SAW signals. The averaged signals were converted to digital data using a data acquisition system (DAQ). While we obtain two orthogonal MIM images referred to as MIM-Re and MIM-Im, only MIM-Re is presented in the present study. These two images have a $\pi$/2 phase shift, but the other feature is typically similar \cite{nii, zheng}. By scanning the tip over the surface of the substrate, we visualized SAW as well as AFM images simultaneously. Since the scanning range of our AFM system is limited to 20 $\mu$m $\times$ 20 $\mu$m, we manually moved the stage using micrometer after obtaining single images. Then, we arranged several images as shown in Fig. 2(c).

\begin{figure}
    \centering
    \includegraphics[width=4.5 in]{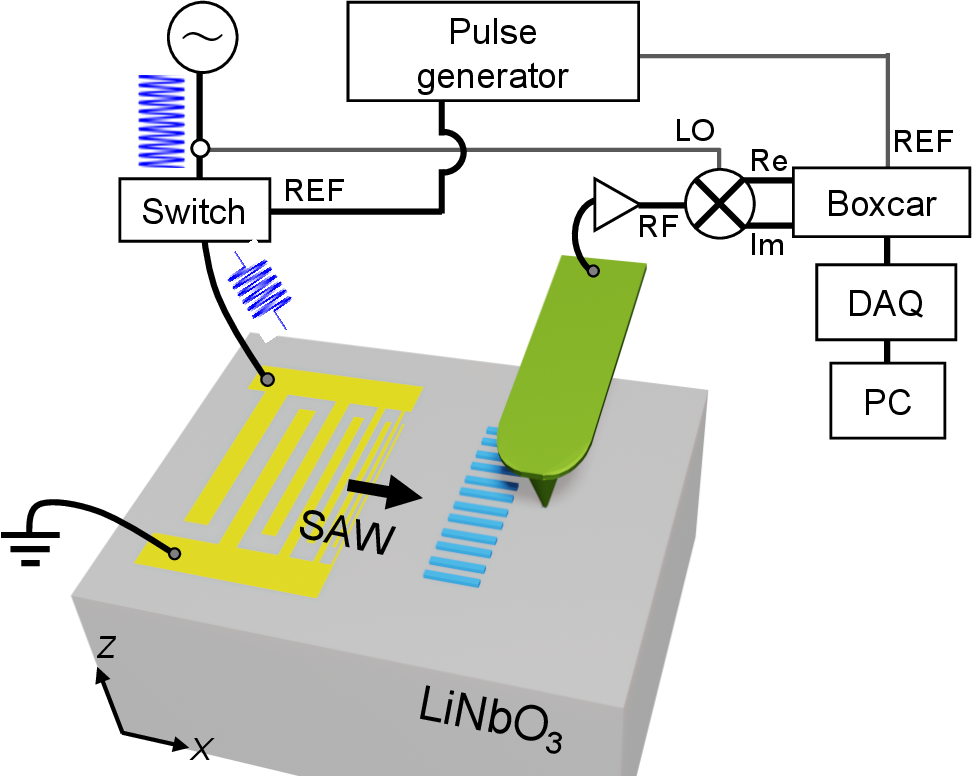}
    \caption{Setup of the MIM experiment. The left IDT excites pulse-modulated SAW, and cantilever tip detects local electric potential induced by SAW. The applied microwave frequency and pulse duration are 2.6 GHz and 100 ns, respectively. The detected microwave signal was demodulated by IQ mixer followed by a time-domain gating by a boxcar integrator. DAQ represents data acquisition system. RF, LO, REF stand for radio-frequency, local oscillator, and reference signals, respectively. Re and Im correspond to in-phase and quadrature-phase signals after demodulation. 
    }
    \label{fig1}
\end{figure}

\subsection{Variation of $\bm{L}_{\rm{SAW}}$ in the momentum space}
To derive the variation of phonon angular momentum $\bm{L}_{\rm{SAW}}$ in the momentum space, we calculated the acoustic eigenmodes of a $Z$-cut LiNbO$_3$ crystal. The calculation is based on a finite element method using a commercial software COMSOL Multiphysics 5.6. As a unit of elastic substance, we employed a rectangular shaped LiNbO$_3$ with the length, width, and thickness of $l_0$ = 3 $\mu$m, $w_0$ = 1 $\mu$m, and $t_0$ = 21 $\mu$m, respectively. We applied Floquet periodic boundary condition to the $x$ and $y$ planes while the $+z$($-z$) planes were treated as free boundary (low-reflecting boundary).\par
Eigenmodes and eigenfrequencies were calculated at a fixed wavevector of ($k_x$, 0, 0), where we chose $k_x$ = 0.7$\times \pi/l_0$ [rad/$\mu$m]. The crystal axes of LiNbO$_3$, denoted as $X_{LN}$, $Y_{LN}$, and $Z_{LN}$, were aligned with the Cartesian coordinate system ($x$, $y$, and $z$). The in-plane anisotropy of LiNbO$_3$ was investigated by rotating the crystal axes around the $z$-axis by an angle $\gamma$. Given that the SAW mode has the lowest sound velocity, our focus was on eigenmodes with the lowest eigenfrequency. The SAW velocity was derived by calculating $v_{SAW} = 2\pi f/k_x$, where $f$ is eigenfrequency.
\par
The angular momentum, denoted as $\bm{L}_{\rm{SAW}}$ = ($L_x$, $L_y$, $L_z$), was calculated using the expression $L_i$ = $\epsilon_{ijk}$ [$\Re[u_k]\Im[u_j]-\Re[u_j]\Im[u_k]$]. Here $i$, $j$, $k$ = $x$ $y$, $z$, and $\bm{u}$ = ($u_x$, $u_y$, $u_z$) is the displacement vector of the obtained eigenmodes with complex amplitude. $\epsilon_{ijk}$ is Levi-Civita symbol. Obtained $\bm{L}_{\rm{SAW}}$ were overlaid on the LiNbO$_3$ structure as color maps [Fig. S5(b) and S5(c)]. In order to evaluate the total angular momentum that can be injected into the Ni film on the surface, we calculated the integral of $\bm{L}_{\rm{SAW}}$ over the surface (+$z$) of LiNbO$_3$. Figure S5(d) shows the surface integrated $\bm{L}_{\rm{SAW}}$ as a function of various crystal rotation ($\gamma$). Since there is a three-fold symmetry about $Z_{LN}$-axis, we show that only from 0 deg to 120 deg. The dominance of $L_y$ at any $\gamma$ is evident, indicating that SAW eigenmode is always Rayleigh-wave-like, irrelevant to $\gamma$. In other words, the total angular momentum points nearly along $y$-direction and orthogonal to both surface normal and wavevector of SAW ($|| x$). At $\gamma$ = 0, 60, 120 deg, an additional component of $L_z$ is present. Consequently, $\bm{L}_{\rm{SAW}}$ is canted along the out-of-plane direction with an angle of $\tan^{-1}(L_z/L_x)$ = $-17$ deg at $\gamma$ = 0 deg and  $+17$ deg at $\gamma$ = 60 deg. The resulting $\bm{L}_{\rm{SAW}}$ direction is schematically displayed in Fig. 1(b) by arrows, along with the corresponding SAW velocity [Fig. 5S(f)] shown as a velocity circle.

\begin{figure}
    \centering
    \includegraphics[width=6.7 in]{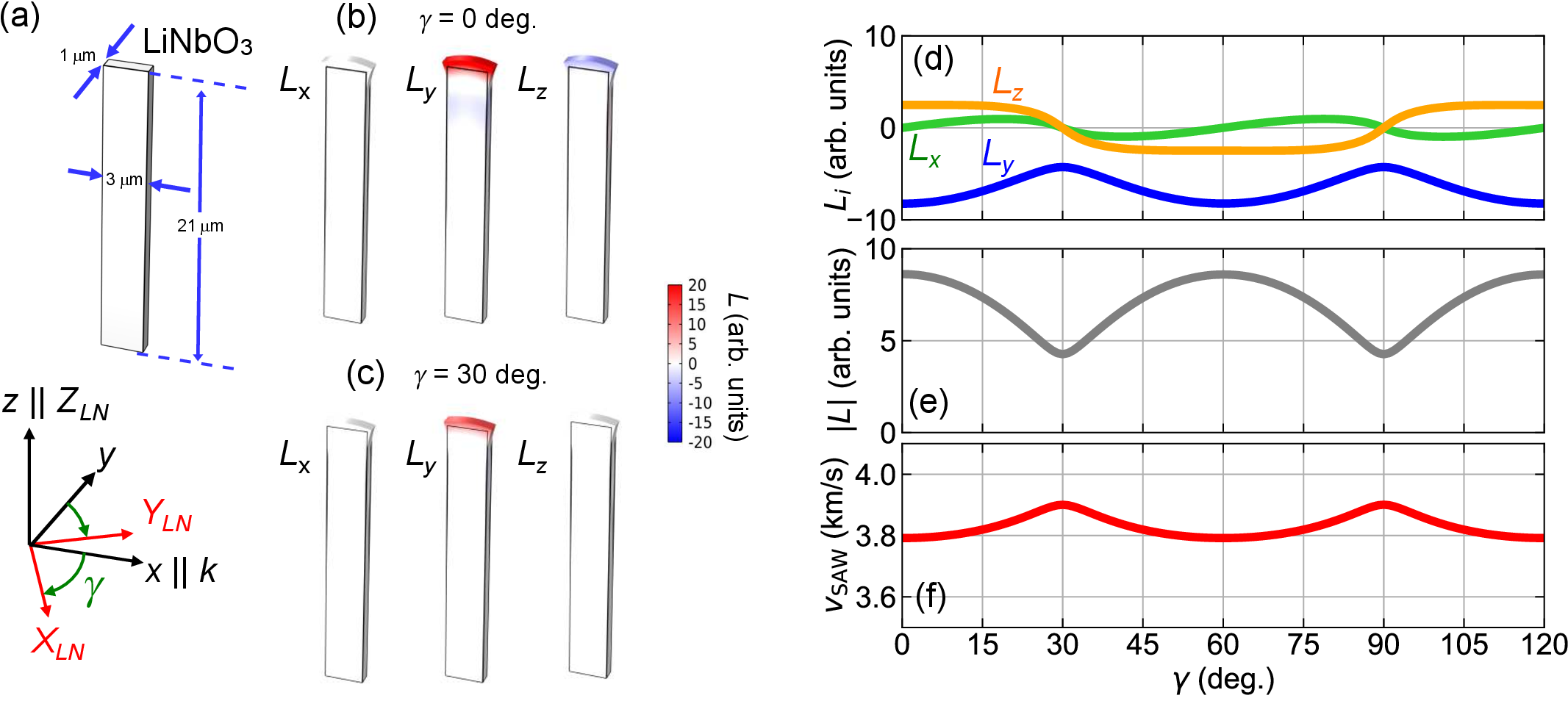}
    \caption{Caculation of $\bm{L}_{\rm{SAW}}$ based on a finite element method. (a) A schematic of the unit structure for the calculation of elastic eigenmodes of LiNbO$_3$. The crystal axes were rotated about $\gamma$ with respect to $z$-axis. (b),(c) $\bm{L}_{\rm{SAW}}$ for SAW eigenmode overlaid onto the unit structure at (b) $\gamma$ = 0 deg and (c) $\gamma$ = 30 deg, respectively. (d)-(f) $\gamma$-dependence of (d) $L_i$ ($i$ = $x$, $y$, $z$) and (e) magnitude of  angular momentum $|\bm{L}|$, and (f) SAW velocity.
    }
    \label{fig1}
\end{figure}

\subsection{Theoretical calculation of scattering amplitude}
\subsubsection{Scattering amplitude}

To interpret the observed nonreciprocal diffraction in terms of a resonant scattering of SAW by magnetic excitations, we model each magnetic wire as a point-like scatterer accompanied by a spin degree of freedom $\bm{S}_{\rm mag}$. Although the interference arising from the periodic array of point scatters is important in amplifying the diffracted signals at the particular angles $\phi = \pm 30$ deg, it is not expected to affect the ratio between the upward and downward diffracted signals \cite{borrmann}. While the Ni wire is longer than the SAW wavelength so that in reality the relevant magnetic excitation would have a spatial profile, it would take a straightforward modification of what follows without anything to gain in interpreting the present experiment. For the sake of brevity, therefore, we focus on the magnetic field dependence of SAW scattering by a single spin and compress any effect of the interference and mode profile into an overall multiplicative factor that cannot be reliably predicted anyway and is to be fitted to the experimental data. 

We choose the coordinate system such that the incident SAW travels along the negative $x$-direction and the positive $z$-direction coincides with the outward normal of the LiNbO$_3$ surface. All the bulk acoustic waves are ignored and the displacement vector $\bm{u}$ is assumed to be expanded in terms of a single mode of SAWs as
\begin{equation}
\bm{u} = \sum _{\bm{q}} \sqrt{\frac{\hbar }{2\rho \omega _{\bm{q}}AL}} \left\{ a_{\bm{q}}\bm{e}_{\bm{q}}\left( z\right) e^{i\bm{q}\cdot \bm{r}} + {\rm h.c.} \right\}  ,
\end{equation}
where $\bm{r} =(x,y)^T$ and $\bm{q} =q( \cos \phi _{\bm{q}}, \sin \phi _{\bm{q}} )^T$ are the in-plane coordinates and wavevector respectively, $A,L$ are the area of the surface and the thickness of the substrate respectively, $\rho $ is the mass density, $\omega _{\bm{q}}= cq $ is the SAW eigenfrequency with $c$ the sound velocity, $a_{\bm{q}}$ is the phonon annihilation operator, and $\bm{e}_{\bm{q}}$ is an appropriately normalized phonon polarization vector which is proportional to $\bm{u}$ in the previous subsection for a fixed $\bm{q}$. If the elastic medium is assumed isotropic with the longitudinal and transverse sound velocities $c_P ,c_S$ respectively and stress-free boundary conditions are imposed, the surface mode should be the Rayleigh waves whose polarization vector is given by
\begin{equation}
\bm{e}_{\bm{q}} = N_R \sqrt{qL} \begin{pmatrix}
	i e_q^{\parallel }\left( z\right) \begin{pmatrix}
	\cos \phi _{\bm{q}} \\
	\sin \phi _{\bm{q}} \\
	\end{pmatrix} \\
	e_q^{\perp }\left( z\right) \\
	\end{pmatrix} ,
\end{equation}
where $N_R $ is the normalization constant of order unity and
\begin{eqnarray*}
e^{\parallel }_q \left( z\right) &=& \left( 1-\frac{c^2}{2c_S^2} \right) \left\{ e^{q\sqrt{1- c^2 / c_P^2 }z} -\left( 1-\frac{c^2}{2c_S^2} \right) e^{q\sqrt{1-c^2/c_S^2}z} \right\}  , \\
e^{\perp }_q \left( z\right) &=& \sqrt{1-\frac{c^2}{c_P^2}}\left\{ \left( 1-\frac{c^2}{2c_S^2} \right) e^{q\sqrt{1-c^2 /c_P^2}z} - e^{q\sqrt{1-c^2 /c_S^2}z} \right\} .
\end{eqnarray*}
Given the actual sample is anisotropic and the surface conditions may not be entirely stress-free, we leave $e_q^{\parallel } ,e_q^{\perp }$ unspecified apart from being real and independent of $\phi _{\bm{q}}$.

We consider the magnetic field lying in-plane $\bm{H} = \left| H \right| (\cos \theta ,\sin \theta ,0)^T$ and take it to be the quantization axis of the spin $\bm{S}_{\rm mag} = (S_x ,S_y ,S_z )^T$. Note that $H<0$ with $0\leq \theta < 180$ deg in the main text is conventionally represented by $\theta \rightarrow \theta + 180$ deg in this section. As usual, the spin states are denoted $|m\rangle , m =-S,-S+1 ,\cdots ,S$ with $S$ the size of the spin, where $( S_x \cos \theta +S_y \sin \theta )|m\rangle = m|m\rangle $ and
\begin{equation}
S_{\pm }|m \rangle = \sqrt{\left( S\mp m\right) \left( S\pm m+1\right)}|m\pm 1\rangle , \quad S_{\pm } = -S_x \sin \theta +S_y \cos \theta \pm iS_z .
\end{equation}
For simplicity, we retain only Zeeman energy in the spin Hamiltonian $\mathcal{H}_S = \mu _B \mu _0 \bm{H}\cdot \bm{S}_{\rm mag}$, where $\mu _B$ and $\mu _0$ are the Bohr magneton and permeability of vacuum respectively. Although there is a significant demagnetizing field of the wire acting on the spins in reality, for our present purposes, it is sufficient to reinterpret $\theta $ to be the angle of the magnetization instead of the magnetic field and appropriately adjust the resonance field, as discussed in the next subsection. The energy levels are then given by $\hbar \omega _m = m\mu _B \mu _0 \left| H \right| $ so that $m=-S$ is the ground state. The total Hilbert space is spanned by states of the form $ \mathcal{N} \prod _{\bm{q}} \left( a_{\bm{q}}^{\dagger }\right) ^{n_{\bm{q}}} |0\rangle \otimes |m\rangle  $ where $n_{\bm{q}}$ is the number of phonons with wavevector $\bm{q}$, $|0\rangle $ is the vacuum state in the phonon Fock space, and $\mathcal{N}$ is the normalization constant. 

We study a SAW scattering process from $|\bm{k}, m,N\rangle $ to $|\bm{q},m, N \rangle $, where $N$ stands for some initial distribution of the phonons $|N\rangle = |\{ \cdots, n_{\bm{k}}, \cdots , n_{\bm{q}}, \cdots \} \rangle $ and $\bm{k}$ and $\bm{q}$ are the incident and scattered phonon wavevectors respectively, i.e. $|\bm{k},m,N\rangle = |\{ \cdots ,n_{\bm{k}}+1 , \cdots ,n_{\bm{q}},\cdots \} \rangle \otimes |m\rangle $ and similarly for $\bm{q}$. Let us consider the total Hamiltonian $\mathcal{H} = \sum _{\bm{q}}\hbar \omega _{\bm{q}}a_{\bm{q}}^{\dagger }a_{\bm{q}}+ \mathcal{H}_S + \mathcal{H}_{\rm me} f\left( t\right) $, where $\mathcal{H}_{\rm me}$ represents the magnetoelastic interactions. According to the elementary quantum mechanics, if the interaction is turned on through some function of time $f\left( t\right) $ that satisfies $f\left( t\right) \rightarrow 0 $ as $t\rightarrow -\infty $, the transition probability from $|\bm{k},m,N\rangle $ to $|\bm{q},m,N\rangle $ in the second order perturbation yields
\begin{equation}
P^{N,m}_{\bm{k}\rightarrow \bm{q}} = \frac{2\pi }{\hbar ^2}\left| M_{\bm{k}\rightarrow \bm{q}}^{N,m} \right| ^2 \left| \tilde{f}\left( \omega _{\bm{q}}-\omega _{\bm{k}} \right) \right| ^2 ,
\end{equation}
where $\tilde{f}$ is the Fourier transform of $f\left( t\right) $. $M^{N,m}_{\bm{k}\rightarrow \bm{q}}$ is the scattering amplitude given by
\begin{equation}
M_{\bm{k}\rightarrow \bm{q}}^{N,m} = \frac{1}{\hbar }\sum _{\alpha } \frac{\langle \bm{q},m,N | \mathcal{H}_{\rm me} |\alpha \rangle \langle \alpha | \mathcal{H}_{\rm me} |\bm{k},m,N\rangle }{\omega _{\alpha } -\omega _{\bm{k}}-\omega _m -\omega _N +i\gamma _{\alpha m} } ,
\end{equation}
where $\alpha $ labels all the possible intermediate states and phenomenological relaxation constants $\gamma _{\alpha m}$ have been introduced by hand. As $\tilde{f}$ appears in an irrelevant multiplicative factor, the problem reduces to the computation of $M_{\bm{k}\rightarrow \bm{q}}^{N,m}$. 

For the magnetoelastic interactions, we take
\begin{equation}
\mathcal{H}_{\rm me} = b \int _V  d^3 r \sum _{i,j=x,y,z} \epsilon _{ij} S_i S_j ,
\end{equation}
where $b$ is a phenomenological magnetoelastic coupling constant, $\epsilon _{ij} = (\partial _j u_i +\partial _i u_j )/2$ are the strain tensor components, and $V$ denotes the spatial volume occupied by the magnetic scatterer. It assumes full rotational symmetry, which is appropriate for polycrystalline Ni wires. The assumption also excludes magneto-rotation coupling~\cite{Maekawa,xu}, which should accompany the shape anisotropy field to be discussed in the next subsection. Nickel has a large magneto-elastic coupling $b$ far exceeding 10$^6$ J/m$^3$~\cite{Lee_1961} and a relatively small saturation magnetization $\mu _0 M_s^2 \approx 1.7 \times 10^5$ J/m$^3$~\cite{Weiler, Dreher}, however, so that the shape-anisotropy-induced magneto-rotation coupling is expected to be subdominant. We also remark that the effect of rotation in the present setup is identical to the contribution from off-diagonal shear components $\epsilon _{xz},\epsilon _{yz}$. Although additional magnetic field angle dependences could arise from more complicated terms with lower symmetries, including them would not help unless the coefficients could be quantified by some independent means. We keep those terms linear in $S_{\pm }$ only, obtaining
\begin{equation}
\mathcal{H}_{\rm me} \approx b\int _V d^3 r \left[ S_+ \left( S_x \cos \theta +S_y \sin \theta +\frac{1}{2} \right) \left\{ -\frac{\epsilon _{xx}-\epsilon _{yy}}{2}\sin 2\theta +\epsilon _{xy}\cos 2\theta -i\left( \epsilon _{zx}\cos \theta +\epsilon _{zy}\sin \theta \right) \right\} +{\rm h.c.} \right] .
\end{equation}
By inspection, the only possible intermediate states $|\alpha \rangle $ for which the matrix elements are nonzero are $|\alpha \rangle = |N\rangle \otimes |m\pm 1\rangle \equiv |m\pm 1,N\rangle , |\{ \cdots ,n_{\bm{k}}+1 ,\cdots ,n_{\bm{q}}+1 , \cdots \} \rangle \otimes |m\pm 1\rangle \equiv |\bm{k},\bm{q},m\pm 1,N\rangle $. It is a straightforward matter to calculate those matrix elements:
\begin{eqnarray}
\langle m\pm 1,N|\mathcal{H}_{\rm me}|\bm{k},m,N\rangle &=& b A_{\rm eff}\left( \bm{k}\right)  \sqrt{\frac{\hbar \left( n_{\bm{k}}+1\right)}{2\rho cA}}\left( m\pm \frac{1}{2} \right) \sqrt{\left( S\mp m\right) \left( S\pm m+1\right)}  \nonumber \\
&& \times  \left\{  C_{\parallel } \left( k \right) \sin \left( \theta -\phi _{\bm{k}}\right) \pm C_{\perp }\left( k \right)   \right\}  \cos \left( \theta -\phi _{\bm{k}}\right) , \\
\langle \bm{k},\bm{q},m\pm 1,N |\mathcal{H}_{\rm me} |\bm{k},m,N\rangle &=& b \overline{A_{\rm eff}\left( \bm{q}\right)}  \sqrt{\frac{\hbar \left( n_{\bm{q}}+1 \right)}{2\rho cA}} \left( m\pm \frac{1}{2} \right) \sqrt{\left( S\mp m\right) \left( S\pm m+1 \right)} \nonumber \\
&& \times \left\{  C_{\parallel }\left( q \right) \sin \left( \theta -\phi _{\bm{q}}\right) \mp C_{\perp }\left( q \right)  \right\} \cos \left( \theta -\phi _{\bm{q}}\right)   ,
\end{eqnarray}
where $A_{\rm eff}\left( \bm{k}\right) = \int _{\rm scatterer} d^2 r  e^{i\bm{k}\cdot \bm{r}}$ is the effective area of the scatterer seen from the traveling SAW, overbar denotes complex conjugation, and appropriately normalized and thickness-averaged longitudinal and transverse components of the strain are given by
\begin{eqnarray}
C_{\parallel }\left( k \right) &=& N_R \int d\left( kz \right) e^{\parallel }_k \left( z\right) ,  \\
C_{\perp }\left( k \right) &=& N_R \int d\left( kz\right) \frac{de^{\parallel }_k \left( z\right) /d\left( kz\right) + e^{\perp }_k \left( z\right)}{2} .
\end{eqnarray}
Setting $\phi _{\bm{k}} = \pi $ and $\phi _{\bm{q}} =\phi +\pi  $, one obtains
\begin{eqnarray}
M_{\bm{k}\rightarrow \bm{q}}^{N,m} &=& \frac{b^2}{2\rho cA}\sqrt{\left( n_{\bm{k}}+1\right) \left( n_{\bm{q}}+1\right) } \overline{A_{\rm eff}\left( q\right)}A_{\rm eff}\left( k\right) \cos \theta \cos \left( \theta -\phi \right) \nonumber \\
&& \times \sum _{\pm }\left( m\pm \frac{1}{2} \right) ^2 \left( S\mp m\right) \left( S\pm m+1\right) \Bigg[ \frac{\left\{ C_{\perp }\left( q\right) \mp  C_{\parallel }\left( q\right) \sin \left( \theta -\phi \right) \right\} \left\{ C_{\perp }\left( k\right) \mp C_{\parallel }\left( k\right) \sin \theta \right\}}{\pm \mu _B \mu _0 \left| H\right| /\hbar -\omega _{\bm{k}}+i\gamma _{m\pm 1,m}} \nonumber \\
&& + \frac{\left\{ C_{\perp }\left( q\right)  \pm  C_{\parallel }\left( q\right) \sin \left( \theta -\phi \right) \right\} \left\{ C_{\perp }\left( k\right) \pm C_{\parallel }\left( k\right) \sin \theta \right\}}{\pm \mu _B \mu _0 \left| H\right| /\hbar +\omega _{\bm{q}}+i\gamma _{m\pm 1,m}} \Bigg] .
\end{eqnarray}
Finally assuming the resonance condition is met $\omega _{\bm{k}}\approx \omega _{\bm{q}} \approx \mu _B \mu _0 \left| H\right| /\hbar $ and dropping the anti-resonant contributions, one derives
\begin{equation}
M_{\bm{k}\rightarrow \bm{q}}^{N,m} \approx C_0 \left\{ C_{\perp }\left( q\right) -  C_{\parallel }\left( q\right) \sin \left( \theta -\phi \right) \right\} \left\{ C_{\perp } \left( k\right) -C_{\parallel }\left( k\right) \sin \theta \right\}  \cos \theta \cos \left( \theta -\phi \right) , \label{eq:amplitude}
\end{equation}
where the field independent factor $C_0$ is given by
\begin{eqnarray}
C_0 &=& \frac{b^2}{2\rho cA}\sqrt{\left( n_{\bm{k}}+1\right) \left( n_{\bm{q}}+1\right)}\overline{A_{\rm eff}\left( q\right)}A_{\rm eff}\left( k\right) \nonumber \\
&& \times  \left\{ \left( m+\frac{1}{2}\right) ^2 \frac{\left( S-m\right) \left( S+m+1\right)}{\mu _B \mu _0 H/\hbar -\omega _{\bm{k}}+i\gamma _{m+1,m}} +\left( m-\frac{1}{2}\right) ^2 \frac{\left( S+m\right) \left( S-m+1\right)}{-\mu _B \mu _0 H/\hbar +\omega _{\bm{q}}+i\gamma _{m-1,m}} \right\} .
\end{eqnarray}

While we have so far discussed scattering by the spin degree of freedom, the magnetic wire also carries mechanical degrees of freedom that also scatter SAW independently of the magnetic field. The total scattering amplitude is given by the sum of the mechanical scattering~\cite{klemens, tamura} and the magnetoelastic one as
\begin{equation}
M_{tot}^{N,m}  = M_0^{N,m} + M_{\bm{k}\rightarrow\bm{q}}^{N,m},
\end{equation}
where $M_0^{N,m} = \braket{\bm{q},m,N|\mathcal{H}_0|\bm{k},m,N}$ is the static scattering amplitude that arises from some mechanical Hamiltonian $\mathcal{H}_0$ representing the wire and independent of the magnetic field. Being mechanical and a linear order contribution, $M_0^{N,m}$ is expected to be much larger than $M_{\bm{k}\rightarrow \bm{q}}^{N,m}$, which is consistent with the large background amplitude in the experiment. Therefore, $\left| M_{\rm tot}^{N,m} \right| ^2 \approx \left| M_0^{N,m} \right| ^2  +2 \Re \left[ \overline{M_0^{N,m}}M_{\bm{k}\rightarrow \bm{q}}^{N,m} \right] $. Noting $M_{\bm{k}\rightarrow \bm{q}}^{N,m} \rightarrow 0$ as $H\rightarrow \infty $, the experimentally defined change of scattering intensity $\Delta I$ with respect to a sufficiently large reference field $H_{\rm ref}$ should be proportional to $ \left| M_{\rm tot}^{N,m} \right| ^2 - \left| M_0^{N,m} \right| ^2 \approx 2\Re \left[ \overline{M_0^{N,m}}M_{\bm{k}\rightarrow \bm{q}}^{N,m} \right] $, averaged over a probability distribution $p_{N,m}$ of $N$ and $m$. However, we are interested in the dependence on the magnetic field angle only, which according to Eq.~(\ref{eq:amplitude}) does not depend on $N,m$. Therefore, the multiplication by $\overline{M_0^{N,m}}$, taking the real part, and averaging over $N,m$ will all do nothing concerning the angular dependence and they can be accounted for in an appropriately redefined overall factor $\tilde{C}_0$. To summarize, we use
\begin{equation}
\frac{\Delta I}{I_0} = \frac{\sum _{N,m} p_{N,m}\left( \left| M_{\rm tot}^{N,m} \right| ^2 -\left| M_0^{N,m} \right| ^2 \right) }{\sum _{N,m} p_{N,m} \left| M_{\rm tot}^{N,m} \right| ^2 } \approx \tilde{C}_0 \left\{ C_{\perp }\left( q\right) - C_{\parallel }\left( q\right) \sin \left( \theta -\phi \right) \right\} \left\{ C_{\perp }\left( k\right) -C_{\parallel } \left( k\right) \sin \theta \right\} \cos \theta \cos \left( \theta -\phi \right) \label{eq:intensity}
\end{equation}
to fit the experimentally observed relative change in the scattering amplitude. We recall that the magnetization direction is not perfectly parallel to the applied field since there is a significant magnetic shape anisotropy in our Ni grating sample. Note that the anisotropy in resonance frequency will not affect the angle dependence as Eq.~(\ref{eq:intensity}) assumes the resonance condition. Thus, we estimated magnetization direction $\theta_0$ at various magnetic field directions $\theta$ as described in the following subsection and performed the fitting using Eq.~(\ref{eq:intensity}) with $\theta $ substituted by $\theta _0$.

Figures S6 compare $\Delta I_{max}/I_0$ for the experiment and calculation.
Here, $\Delta I_{max}/I_0 = (I_{max} - I_0)/I_0$ is plotted for the experiment, where we consider the maximal relative change in the intensity at positive ($H >$ 0) or negative ($H <$ 0) magnetic fields, respectively. Corresponding to this, the field dependent parameter $\tilde{C}_0$ is regarded as a constant. The best fitting parameters are $\tilde{C}_0 =  -5.00$, $C_{\perp }(q) = 0.478$, $C_{\parallel }(q) = 0.147$, $C_{\perp }(k) = 0.769$, and $C_{\parallel }(k) = -0.070$. We multiplied a factor of 0.59 to the fitting curves at $\phi = -30$ deg. This factor may be attributable to variations in the performance between the top ($\phi = 30$ deg) and the bottom ($\phi = -30$ deg) IDTs. The calculated results roughly reproduce experimental features: As shown in Fig. S6(a) and Fig. S6(c),  $\Delta I_{max}/I_0$ is larger for $H>0$ than for $H<0$ when $\theta < 90$ deg, while this qualitative trend reverses when $\theta > 90$ deg. Moreover, these relations are inverted at  $\phi$ = $-$30 deg [Fig. S6(b) and S6(d)]. Nonreciprocity shown in Fig. 4(c) in the maintext was also deduced by subtracting the calculated results for $H>0$ from those for $H<0$. This also shows a good agreement with the experimental angular dependence.

As a closing remark, we note that the fitted parameters imply that the transverse strain component $\epsilon _{xz}\cos \phi _{\bm{k}} + \epsilon _{yz}\sin \phi _{\bm{k}} $ dominates over the longitudinal ones $\epsilon _{xx} , \epsilon _{yy} ,$ and horizontal transverse one $\epsilon _{xy}$. This feature is also confirmed by the magnetic field angle dependence of SAW absorption as shown in Fig. S7. The fitting result gives transverse shear strain $\epsilon_{xz}$ 10 times larger than the longitudinal one $\epsilon_{xx}$. This is in contrast to the prediction from the usual SAW theory that assumes a stress-free boundary condition, which enforces $\epsilon _{xz} =\epsilon _{yz} =0 $ at the surface and suppresses their contributions for films thinner than the SAW wavelength. Taking account of the magneto-rotation coupling would not change the situation qualitatively. While the disagreement might be due to the surface strains or a modified thickness profile of the SAW in the real samples, modelling them would require a much more detailed knowledge of the experimental conditions. Another possible source of discrepancy is the anisotropy of the magneto-elastic interaction, which was ignored in our simple model, although including these features in general would lead to too many fitting parameters. It therefore requires further investigations and is left for future studies.

\begin{figure}
    \centering
    \includegraphics[width=4.5in]{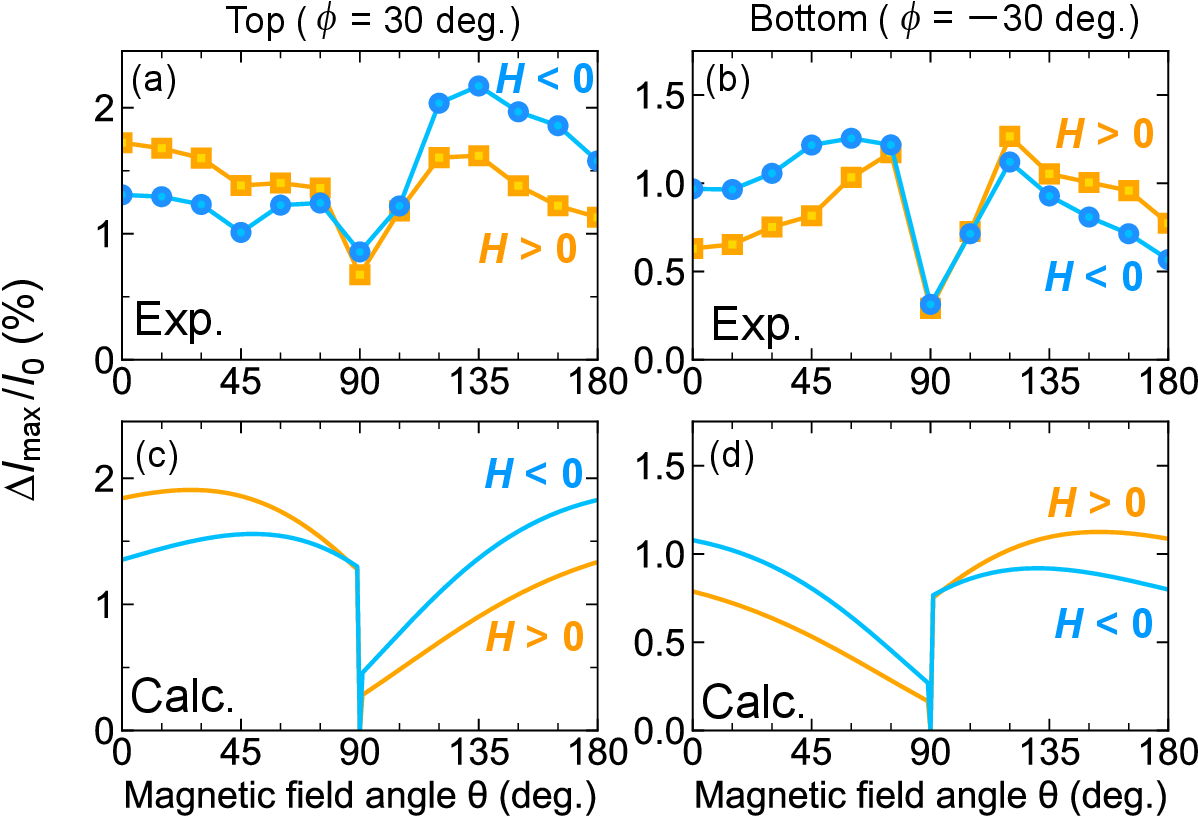}
    \caption{Magnetic field angle dependence of $\Delta I_{max}/I_0$ measured by (a) top and (b) bottom IDTs. (c), (d) Corresponding calculation of $\Delta I_{max}/I_0$ for (c) top and (d) bottom IDTs, respectively.
    }
    \label{fig1}
\end{figure}

\begin{figure}
    \centering
    \includegraphics[width = 6 in]{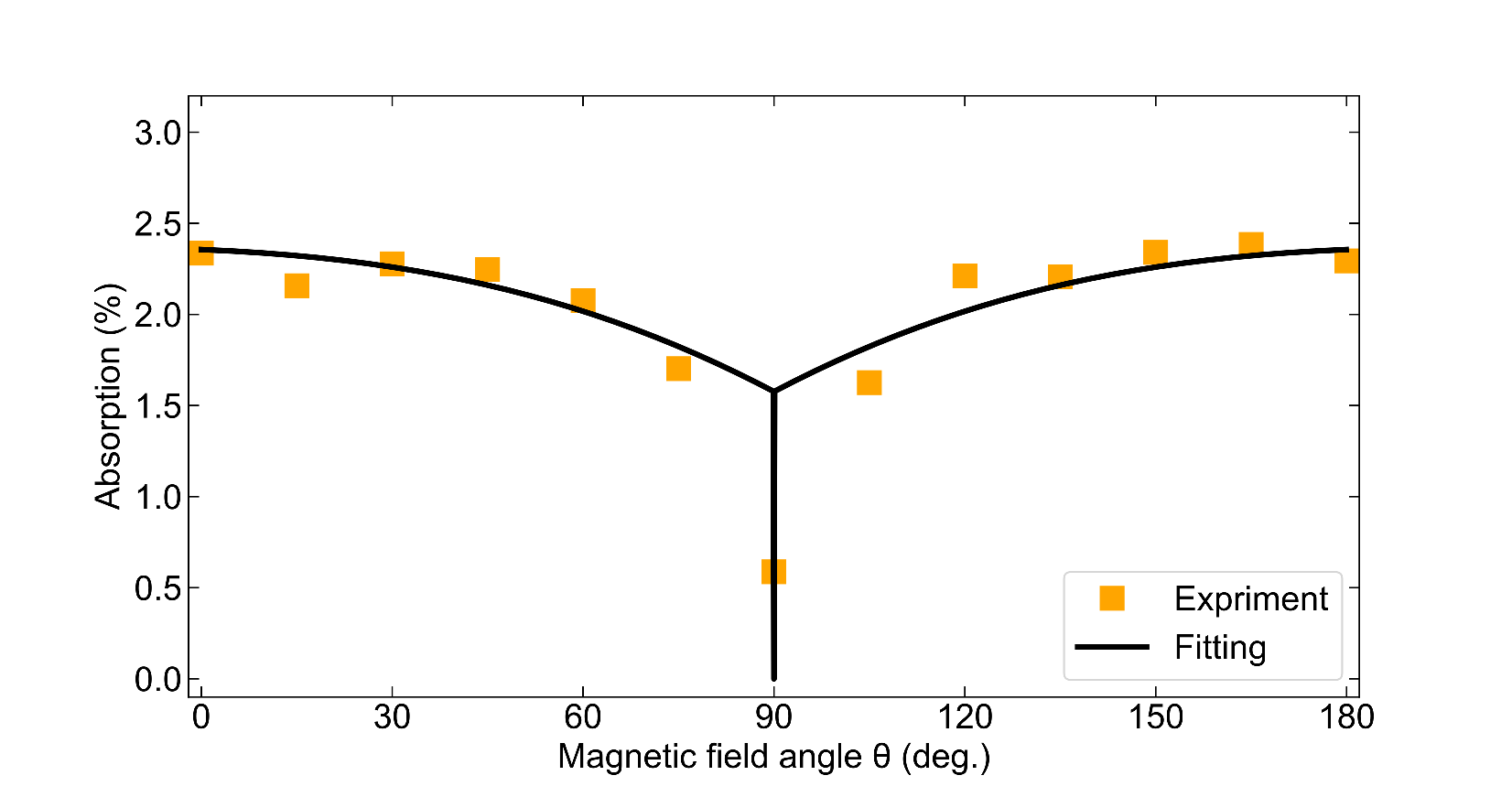}
    \caption{Magnetic field angle dependence of SAW absorption. The maximal absorption was deduced from $\Delta I/I_0$ of transmitted SAW intensity obtained from the center IDT. The black solid line is fitting considering a conventional acoustic-FMR induced by Rayleigh-type SAW. Assuming the isotropic magnetoelastic coupling $b_1 = b_2$,  the SAW absorption is given by $\propto |\epsilon_{xx}\cos\theta_0 \sin \theta_0 + \epsilon_{zx}\cos \theta_0|^2$ \cite{Dreher, sasaki}. Here, $\epsilon_{xx}$ and $\epsilon_{zx}$ represent longitudinal and transverse shear strains, respectively. Fitting by this equation gives $|\epsilon_{xx}| : |\epsilon_{zx}| = 1.0 : 10.3$, indicating the shear strain plays a dominant role in our device. In this fitting, we used the angular relation between $\bm{M}$ and $\bm{H}$ shown in Fig. S8(c).}
    \label{fig1}
\end{figure}

\subsubsection{Derivation of magnetization direction}
The Ni grating consists of nanowires, which is expected to have a large easy-axis-type magnetic anisotropy. Therefore, we need to deduce magnetization direction ($\theta_0$) in an applied magnetic field with the angle  $\theta$. Here these angles are defined as shown in the inset of Fig. S8(c). To derive magnetization angle, we considered the free energy consisting of Zeeman energy ($E_Z$) and magnetic anisotropy energy ($E_A$) as
\begin{align}
E&= E_Z + E_A  \nonumber \\
&= -MH\cos(\theta-\theta_0)+K_u M_s^2 \sin^2\theta_0,
\end{align}
where $M$, $M_s$, $H$, $K_u$ are magnetization, saturated magnetization, external magnetic field, and magnetic anisotropy constant, respectively.
One can derive stable magnetization direction $\theta_0 = \theta_{0,min}$ by imposing a minimization condition $\partial 
E/\partial \theta_0 = 0$, and $\partial^2 E/\partial \theta_0^2 > 0$. The partial derivatives of the total energy with respect to $\theta_0$ give
\begin{align}
&\displaystyle\frac{\partial E}{\partial \theta_0}
= -HM\sin(\theta-\theta_0)+2 K_u M_s^2 \sin\theta_0 \cos\theta_0,\\
&\displaystyle\frac{\partial^2 E}{\partial \theta_0^2}
= MH\cos(\theta-\theta_0)+K_u M_s^2\cos2\theta_0.
\end{align}
$K_u$ is determined as 0.95$\times 10^{-7}$ [N/$\rm{A}^2$] so that the experimental coercive field $H_c$ is reproduced. We experimentally estimated $H_c$ by using the magnetic field dependence of $\Delta I/I_0$ measured by the center IDT [Fig. S8(a)]. The theoretical $H_c$ can be obtained as the sign reversal field of $\partial^2 E/\partial \theta_0^2$ at the extreme value of $E$. As shown in Fig. S8(b), the calculated $H_c$ successfully reproduced the experimental value.
\begin{figure}
    \centering
    \includegraphics[width=6.375in]{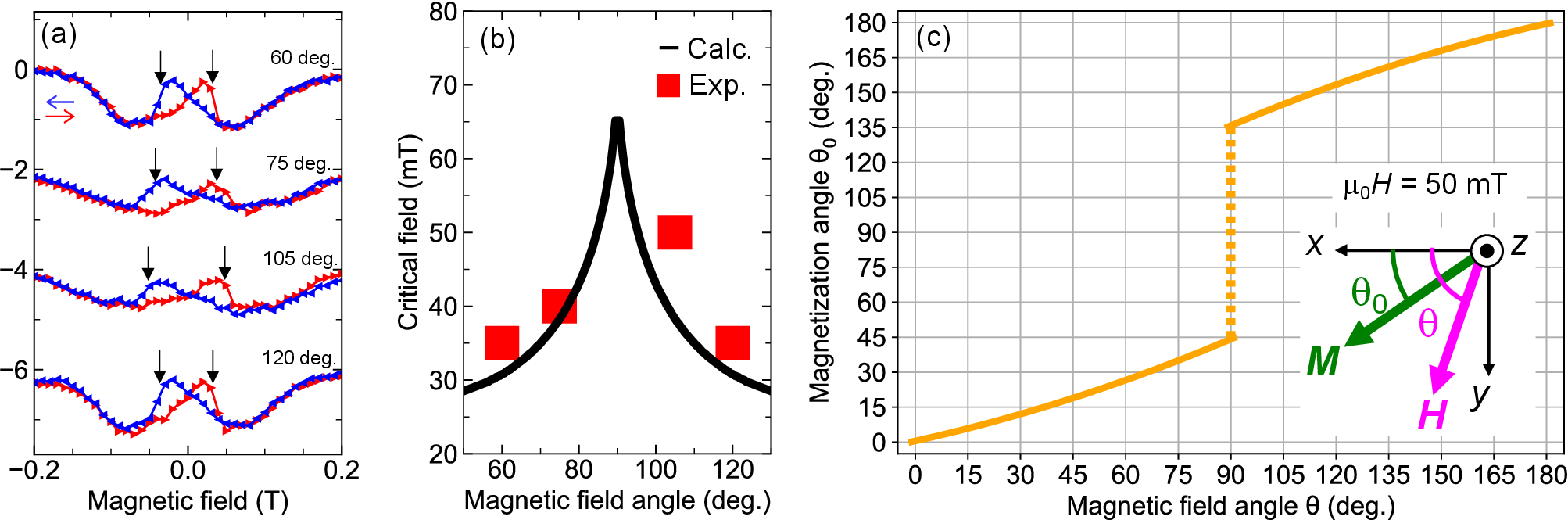}
    \caption{Derivation of angular relationship between magnetization and magnetic field. (a) Magnetic field dependence of transmitted $\Delta I/I_0$ measured by the center IDT at several magnetic field angles $\theta$. The black arrows correspond to a  jump of $\Delta I/I_0$ that indicates to the critical field for magnetization flip. (b) The critical magnetic field for magnetization reversal as a function of the field angle. The solid line shows theoretical fitting. Here we assumed $M = M_s = 370 $ [kA/m] \cite{Weiler, Dreher} and obtained $K_u$ = of 0.95$\times 10^{-7}$ [N/$\rm{A}^2$]. (c) The angular relationship between $\bm{M}$ and $\bm{H}$ using the obtained value of $K_u$. The Ni nanowires are on the $Z$-plane and its longitudinal direction is parallel to $x$-direction. Here the magnetic field of $\mu_0 H$ = 50 mT is considered.
    }
    \label{fig1}
\end{figure}

\bibliography{SAW_supple}